\documentclass{ws-p8-50x6-00}

\newcommand{\beq}{\begin{equation}}
\newcommand{\eeq}{\end{equation}}
\newcommand{\ba}{\begin{array}}
\newcommand{\ea}{\end{array}}

\newcommand{\bef}{\begin{figure}}
\newcommand{\eef}{\end{figure}}
\newcommand{\bce}{\begin{center}}
\newcommand{\ece}{\end{center}}

\newcommand {\Dslash}{D\!\!\!/}

\newcommand{\gsim}{\mathrel{\hbox{\rlap{\lower.55ex \hbox {$\sim$}}
                   \kern-.3em \raise.4ex \hbox{$>$}}}}
\newcommand{\lsim}{\mathrel{\hbox{\rlap{\lower.55ex \hbox {$\sim$}}
                   \kern-.3em \raise.4ex \hbox{$<$}}}}

\begin{document}
\title{Strange Goings on in Quark Matter}
\author{Thomas Sch\"afer}
\address{ Department of Physics and Astronomy, 
     State University of New York, 
     Stony Brook, NY 11794-3800 
     and Riken-BNL Research Center, Brookhaven National 
     Laboratory, Upton, NY 11973
}

\maketitle              

\abstracts{
 We review recent work on how the superfluid state of 
three flavor quark matter is affected by non-zero quark
masses and chemical potentials. }

\section{Introduction}
\label{sec_intro}

  The study of hadronic matter at high baryon density has 
recently attracted a lot of interest. At zero baryon density
chiral symmetry is broken by a quark-anti-quark condensate. 
At high density condensation in the quark-anti-quark channel
is suppressed. Instead, attractive interactions in the 
color anti-symmetric quark-quark channel favor the formation
of diquark condensates. As a consequence, cold dense 
quark matter is expected to be a color 
superconductor\cite{Bailin:1984bm,Alford:1998zt,Rapp:1998zu}. 
The symmetry breaking pattern depends on the density, the 
number of quark flavors, and their masses. A particularly 
symmetric phase is the color-flavor-locked (CFL) phase of three 
flavor quark matter\cite{Alford:1999mk}. This phase is believed 
to be the true ground state of ordinary matter at very large 
density\cite{Schafer:1999ef,Schafer:1999fe,Evans:2000at}. 

  The CFL phase is characterized by the order parameter
\be
\label{cfl}
\langle q_{L,i}^a C q_{L,j}^b\rangle
 = -\langle q_{R,i}^a C q_{R,j}^b\rangle
 = \phi \left(\delta_i^a\delta_j^b-\delta_i^b\delta_j^a\right).
\ee
This order parameter breaks both the global $SU(3)_L\times SU(3)_R
\times U(1)_V$ flavor symmetry and the local $SU(3)_C$ color symmetry
of QCD. As a result, all fermions are gapped and all gluons acquire
a mass via the Higgs mechanism. Color-flavor-locking leaves a 
vectorial $SU(3)_V$ unbroken. This symmetry is the diagonal
subgroup of the original $SU(3)_L\times SU(3)_R\times SU(3)_C$
symmetry. This means that the  color-flavor-locked phase 
exhibits the chiral symmetry breaking patters $SU(3)_L\times 
SU(3)_R\to SU(3)_V$, just like QCD at zero baryon
density. However, the mechanism of chiral symmetry breaking
is quite unusual. The primary order parameter (\ref{cfl})
does not couple left and right-handed quarks. Chiral symmetry
is broken because both left and right-handed flavor are
``locked'' to color, and because color is a vectorial
symmetry. 

  At baryon densities relevant to astrophysical objects dis\-tor\-tions 
of the pure CFL state due to non-zero quark masses are probably 
important\cite{Alford:1999pa,Schafer:1999pb,Schafer:2000ew,Bedaque:2001je,Kaplan:2001qk}.
In the present work we wish to study this problem using 
the effective chiral theory of the CFL phase\cite{Casalbuoni:1999wu}
(CFL$\chi$Th).

\section{CFL Chiral Theory (CFL$\chi$Th)}
\label{sec_CFLchi}

 For excitation energies smaller than the gap the only 
relevant degrees of freedom are the Goldstone modes 
associated with the breaking of chiral symmetry and
baryon number. The interaction of the Goldstone modes
is described by an effective Lagrangian of the 
form\cite{Casalbuoni:1999wu}
\bea
\label{l_cheft}
{\cal L}_{eff} &=& \frac{f_\pi^2}{4} {\rm Tr}\left[
 \nabla_0\Sigma\nabla_0\Sigma^\dagger - v_\pi^2
 \partial_i\Sigma\partial_i\Sigma^\dagger \right] 
 + \Big[ A_1{\rm Tr}(M\Sigma^\dagger){\rm Tr} (M\Sigma^\dagger) \\
 & & \mbox{}
          + A_2{\rm Tr}(M\Sigma^\dagger M\Sigma^\dagger)   
          + A_3{\rm Tr}(M\Sigma^\dagger){\rm Tr} (M^\dagger\Sigma)
         + h.c. \Big]+\ldots . 
 \nonumber 
\eea
Here $\Sigma=\exp(i\phi^a\lambda^a/f_\pi)$ is the chiral field
and $f_\pi$ is the pion decay constant. We have suppressed the 
singlet fields associated with the breaking of the exact $U(1)_V$ 
and approximate $U(1)_A$ symmetries. The theory (\ref{l_cheft})
looks superficially like ordinary chiral perturbation theory.
There are, however, some important differences. Lorentz 
invariance is broken and Goldstone modes travel with the 
velocity $v_\pi<c$. In the CFL phase the ordinary chiral 
condensate $\langle\bar{\psi}\psi\rangle$ is small and
the dominant order parameter for chiral symmetry breaking
is $\langle(\bar{\psi}\psi)^2\rangle$. As a consequence,
the coefficient of ${\rm Tr}(M\Sigma)$ is exponentially
small and the leading mass terms are quadratic in $M$. 

 The pion decay constant $f_\pi$ and the coefficients 
$A_{1,2,3}$ can be determined using matching techniques. 
Matching expresses the requirement that Green functions 
in the effective chiral theory and the underlying microscopic 
theory, QCD, agree. The pion decay constant is most easily 
determined by coupling gauge fields $W_{L,R}$ to the left 
and right flavor currents. As usual, this amounts to 
replacing ordinary derivatives by covariant derivatives. 
The time component of the covariant derivative is given by 
$\nabla_0\Sigma=\partial_0 \Sigma+iW_L\Sigma-i\Sigma W_R$ 
where we have suppressed the vector index of the gauge fields. 
In the CFL vacuum $\Sigma=1$ the axial gauge field $W_L-W_R$ 
acquires a mass by the Higgs mechanism. From (\ref{l_cheft}) 
we get
\be
\label{wm2}
{\cal L} = \frac{f_\pi^2}{4} \, \frac{1}{2} (W_L-W_R)^2.
\ee
The coefficients $A_{1,2,3}$ can be determined by computing
the shift in the vacuum energy due to non-zero quark masses
in both the chiral theory and the microscopic theory. In the 
chiral theory we have 
\be 
\Delta{\cal E}=  
 -\Big[ A_1\left({\rm Tr}(M)\right)^2
      + A_2{\rm Tr}(M^2) + A_3{\rm Tr}(M){\rm Tr} (M^\dagger)
         + h.c. \Big].
\ee

\section{High Density Effective Theory (HDET)}
\label{sec_hdet}

 In this section we shall determine the mass of the gauge field
and the shift in the vacuum energy in the CFL phase of QCD
at large baryon density. This is possible because asymptotic
freedom guarantees that the effective coupling is weak. The 
QCD Lagrangian in the presence of a chemical potential
is given by
\be
\label{qcd}
 {\cal L} = \bar\psi \left( i\Dslash +\mu\gamma_0 \right)\psi
 -\bar\psi_L M\psi_R - \bar\psi_R M^\dagger \psi_L 
 -\frac{1}{4}G^a_{\mu\nu}G^a_{\mu\nu},
\ee
where $M$ is a complex quark mass matrix which transforms as
$M\to LMR^\dagger$ under chiral transformations $(L,R)\in
SU(3)_L\times SU(3)_R$ and $\mu$ is the baryon chemical potential.
If the baryon density is very large perturbative QCD calculations
can be further simplified. The main observation is that the 
relevant degrees of freedom are particle and hole excitations 
in the vicinity of the Fermi surface. We shall describe these 
excitations in terms of the field $\psi_+(\vec{v}_F,x)$, where 
$\vec{v}_F$ is the Fermi velocity. At tree level, the quark field 
$\psi$ can be decomposed as $\psi=\psi_++\psi_-$ where $\psi_\pm
=\frac{1}{2}(1\pm\vec{\alpha}\cdot\hat{v}_F)\psi$. Integrating 
out the $\psi_-$ field at leading order in $1/p_F$ we 
get\cite{Hong:2000tn,Hong:2000ru,Beane:2000ms}
\bea
\label{fs_eff}
{\cal L} &=& 
 \psi_{L+}^\dagger (iv\cdot D) \psi_{L+}
  - \frac{ \Delta}{2}\left(\psi_{L+}^{ai} C \psi_{L+}^{bj}
 \left(\delta_{ai}\delta_{bj}-
           \delta_{aj}\delta_{bi} \right) 
           + {\rm h.c.} \right) \nonumber \\ 
& & \hspace{0.5cm}\mbox{}
  - \frac{1}{2p_F} \psi_{L+}^\dagger \left(  (\Dslash_\perp)^2 
  + MM^\dagger \right)  \psi_{L+}  
  + \left( R\leftrightarrow L, M\leftrightarrow M^\dagger \right)  + \ldots ,
\eea
where $D_\mu=\partial_\mu+igA_\mu$, $v_\mu=(1,\vec{v})$ and
 $i,j,\ldots$ and $a,b,\ldots$ denote flavor and color 
indices. The longitudinal and transverse components of 
$\gamma_\mu$ are defined by $(\gamma_0,\vec{\gamma})_{\|}=
(\gamma_0,\vec{v}(\vec{\gamma}\cdot\vec{v}))$ and $(\gamma_\mu)_\perp 
= \gamma_\mu-(\gamma_\mu)_{\|}$. In order to perform perturbative 
calculations in the superconducting phase we have added a 
tree level gap term $\psi_{L,R} C\Delta \psi_{L,R}$. 

  The mass of a flavor gauge field can be determined by 
computing the corresponding polarization function in the limit
$q_0=0$, $\vec{q}\to 0$. We find $\Pi^{LL}_{00}=\Pi^{RR}_{00}=
-\Pi^{LR}_{00}=m_D^2/4$ with $m_D^2=(21-8\log(2))p_F^2/(36\pi^2)$.
Matching against equ.~(\ref{wm2}) we get\cite{Son:1999cm} 
\be
f_\pi^2 = \frac{21-8\log(2)}{18} 
  \left(\frac{p_F^2}{2\pi^2} \right) .
\ee 
Our next task is to compute the mass dependence of the vacuum 
energy. To leading order in $1/p_F$ there is only one 
operator in the high density effective theory
\be 
\label{kin}
{\cal L} = -\frac{1}{2p_F} \left( \psi_{L+}^\dagger MM^\dagger \psi_{L+}
 + \psi_{R+}^\dagger M^\dagger M\psi_{R+} \right).
\ee
This term arises from expanding the kinetic energy of a massive
fermion around $p=p_F$. We note that $MM^\dagger/(2p_F)$ and
$M^\dagger M/(2p_F)$ act like effective chemical potentials 
for left and right-handed fermions, respectively. Indeed, to 
leading order in the $1/p_F$ expansion, the Lagrangian (\ref{fs_eff}) 
is invariant under a time dependent flavor symmetry $\psi_{L} 
\to L(t)\psi_{L}$, $\psi_{R}\to R(t)\psi_{R}$ where $X_L=
MM^\dagger/(2p_F)$ and $X_R=M^\dagger M/(2p_F)$ transform
as left and right-handed flavor gauge fields. If we impose 
this approximate gauge symmetry on the CFL chiral theory we 
have to include the effective chemical potentials 
$X_{L,R}$ in the covariant derivative of the chiral field, 
\be
\label{mueff}
 \nabla_0\Sigma = \partial_0 \Sigma 
 + i \left(\frac{M M^\dagger}{2p_F}\right)\Sigma
 - i \Sigma\left(\frac{ M^\dagger M}{2p_F}\right) .
\ee
$X_L$ and $X_R$ contribute to the vacuum energy at $O(M^4)$
\be
\label{E_m4}
\Delta {\cal E} = \frac{f_\pi^2}{8p_F^2} 
 {\rm Tr}\left[(MM^\dagger)(M^\dagger M)-(MM^\dagger)^2\right].
\ee
This result can also be derived directly in the microscopic 
theory\cite{Bedaque:2001je}. This means that we do not have to 
rely on the effective gauge symmetry in order 
to derive (\ref{mueff}). $O(M^2)$ terms in the vacuum energy are
generated by terms in the high density effective theory that
are higher order in the $1/p_F$ expansion. We recently argued
that these terms can be determined by computing chirality 
violating quark-quark scattering amplitudes in 
QCD\cite{Schafer:2001za}. At leading order in the $1/p_F$
expansion the chirality violating scattering amplitude can 
be represented as an effective four-fermion operator
\bea
\label{hdet_m}
 {\cal L} &=& \frac{g^2}{8p_F^4}
 \left( ({\psi^A_L}^\dagger C{\psi^B_L}^\dagger)
        (\psi^C_R C \psi^D_R) \Gamma^{ABCD} +
        ({\psi^A_L}^\dagger \psi^B_L) 
        ({\psi^C_R}^\dagger \psi^D_R) \tilde{\Gamma}^{ACBD} \right)
        \nonumber \\ 
 & & \mbox{} 
 + \Big(L\leftrightarrow R, M\leftrightarrow M^\dagger \Big) . 
\eea
Here, we have introduced the CFL eigenstates $\psi^A$ defined 
by $\psi^a_i=\psi^A (\lambda^A)_{ai}/\sqrt{2}$, $A=0,\ldots,8$.
The tensors $\Gamma$ is defined by
\bea 
 \Gamma^{ABCD} &=& \frac{1}{8}\Big\{ {\rm Tr} \left[ 
    \lambda^A M(\lambda^D)^T \lambda^B M (\lambda^C)^T\right] \\
 & & \hspace{2cm}\mbox{}
 -\frac{1}{3} {\rm Tr} \left[
    \lambda^A M(\lambda^D)^T \right]
    {\rm Tr} \left[
    \lambda^B M (\lambda^C)^T\right] \Big\}. \nonumber
\eea
The tensor $\tilde{\Gamma}$ involves both $M$ and $M^\dagger$
and only contributes to field independent terms ${\rm Tr}
[MM^\dagger]$ in the vacuum energy. We can now compute the 
shift in the vacuum energy due to the effective vertex
(\ref{hdet_m}). The result
\be
\label{E_MM}
\Delta {\cal E} = -\frac{3\Delta^2}{4\pi^2} 
 \left\{  \Big( {\rm Tr}[M]\Big)^2 -{\rm Tr}\Big[ M^2\Big]
   \right\}
 + \Big(M\leftrightarrow M^\dagger \Big)
\ee
determines the coefficients $A_{1,2,3}$ in the CFL
chiral theory. We find 
\be
 A_1= -A_2 = \frac{3\Delta^2}{4\pi^2}, 
\hspace{1cm} A_3 = 0,
\ee
which agrees with the result of Son and Stephanov\cite{Son:1999cm}.

\section{Kaon Condensation}
\label{sec_kcond}

 Using the results discussed in the previous section we
can compute the masses of Goldstone bosons in the CFL phase. 
We have argued that the expansion parameter in the chiral expansion 
of the Goldstone boson masses is\cite{Bedaque:2001je}
$\delta=m^2/(p_F\Delta)$. The first term in this expansion 
comes from the $O(M^2)$ term in (\ref{l_cheft}), but the 
coefficients $A$ contain the additional small parameter 
$\epsilon=(\Delta/p_F)$. In a combined expansion in 
$\delta$ and $\epsilon$ the $O(\epsilon\delta)$ mass term 
and the $O(\delta^2)$ chemical potential term appear at 
the same order. At this order, the masses of the flavored 
Goldstone bosons are
\bea 
\label{mgb}
 m_{\pi^\pm} &=&  \mp\frac{m_d^2-m_u^2}{2p_F} +
         \left[\frac{4A}{f_\pi^2}(m_u+m_d)m_s\right]^{1/2},\nonumber \\
 m_{K_\pm}   &=&  \mp \frac{m_s^2-m_u^2}{2p_F} + 
         \left[\frac{4A}{f_\pi^2}m_d (m_u+m_s)\right]^{1/2}, \\
 m_{K^0,\bar{K}^0} &=&  \mp \frac{m_s^2-m_d^2}{2p_F} + 
         \left[\frac{4A}{f_\pi^2}m_u (m_d+m_s)\right]^{1/2}.\nonumber
\eea
We observe that the pion masses are not strongly affected 
by the effective chemical potential but the masses of the 
$K^+$ and $K^0$ are substantially lowered while the $K^-$ 
and $\bar{K}^0$ are pushed up. As a result the $K^+$ and 
$K^0$ meson become massless if $m_s\sim m_{u,d}^{1/3}\Delta^{2/3}$.
For larger values of $m_s$ the kaon modes are unstable, signaling 
the formation of a kaon condensate. 

 Once kaon condensation occurs the ground state is reorganized.
For simplicity, we consider the case of exact isospin symmetry
$m_u=m_d\equiv m$. The most general ansatz for a kaon condensed 
ground state is given by
\bea
\label{k0+_cond}
\Sigma &=& \exp\left(i\alpha \left[
  \cos(\theta_1)\lambda_4+\sin(\theta_1)\cos(\theta_2)\lambda_5
  \right.\right.\nonumber \\
 & & \hspace{1.5cm}\left.\left.\mbox{}
   + \sin(\theta_1)\sin(\theta_2)\cos(\phi)\lambda_6
   + \sin(\theta_1)\sin(\theta_2)\sin(\phi)\lambda_7
   \right]\right).
\eea
With this ansatz the vacuum energy is given by
\be 
\label{k0+_V}
 V(\alpha) = -f_\pi^2 \left( \frac{1}{2}\left(\frac{m_s^2-m^2}{2p_F}
   \right)^2\sin(\alpha)^2 + (m_{K}^0)^2(\cos(\alpha)-1)
   \right),
\ee
where $(m_K^0)^2= (4A/f_\pi^2)m_{u,d} (m_{u,d}+m_s)$ is the $O(M^2)$ 
kaon mass in the limit of exact isospin symmetry. Minimizing the vacuum 
energy we obtain $\alpha=0$ if $m_s^2/(2p_F)<m_K^0$ and $\cos(\alpha)
=(m_K^0)^2/\mu_{eff}^2$ with $\mu_{eff}=m_s^2/(2p_F)$ if $\mu_{eff}
>m_K^0$. We observe that the vacuum energy is independent of 
$\theta_1,\theta_2,\phi$. The hypercharge density is given by
\be 
n_Y = f_\pi^2 \mu_{eff} \left( 1 -\frac{m_K^4}{\mu_{eff}^4}\right),
\ee
where $\mu_{eff}=m_s^2/(2p_F)$. We observe that within the range 
of validity of the effective theory, $\mu_{eff}<\Delta$, the hypercharge 
density satisfies $n_Y<\Delta p_F^2/(2\pi)$. The upper bound on the 
hypercharge density in the condensate is equal to the particle density 
contained within a strip of width $\Delta$ around the Fermi surface. 

 The symmetry breaking pattern is $SU(2)_I\times U(1)_Y\to 
U(1)$ where $I$ is isospin and $Y$ is hypercharge. This 
corresponds to three broken generators. However, there are 
only two Goldstone modes, the $K^0$ and the $K^+$. This mismatch 
is related to the dispersion relations of the Goldstone 
modes\cite{Miransky:2001tw,Schafer:2001bq}. The $K^0$ is a 
standard Goldstone mode with $\omega\sim p$ whereas the $K^+$ is an
anomalous mode with $\omega\sim p^2$. As explained in more detail 
in\cite{Schafer:2001bq} the appearance of an anomalous Goldstone
mode is related to the fact that the kaon condensed groundstate
has a non-zero expectation value of isospin.

\section{Summary}

 We have studied the groundstate of CFL quark matter for non-zero
quark masses. We have argued that there is a new scale $m_s^2/(2p_F)
\sim \sqrt{m_{u,d}m_s}(\Delta/p_F)$ which corresponds to the onset 
of kaon condensation. For $m_s^2/(2p_F)\sim 1$ CFL pairing breaks
down completely. These results can be established using just
dimensional analysis. If perturbation theory is reliable we can 
be more quantitative. To leading order in $g$, the critical strange 
quark mass for kaon condensation is 
\be
\left. m_s \right|_{crit}= 3.03\cdot  m_d^{1/3}\Delta^{2/3} .
\ee
This result suggests that for values of the strange quark mass 
and the gap that are relevant to compact stars CFL matter is 
likely to support a kaon condensate. 

 It is instructive to compare kaon condensation in CFL matter 
with the kaon condensate discussed by Kaplan and Nelson\cite{Kaplan:1986}. 
Kaplan and Nelson suggested kaon condensation as a path from ordinary 
hadronic matter, which has a deficit of strange quarks, to strange
quark matter. Ideal CFL matter has exactly equal numbers of up, down 
and strange quarks\cite{Rajagopal:2001ff}. However, if the strange 
quark mass is non-zero then ideal CFL matter is oversaturated in 
strangeness. CFL kaon condensation is a way for CFL matter to reduce 
its strangeness content.

 Acknowledgements: I am grateful to P.~Bedaque, D.~T.~Son, 
M.~A. Stephanov, D.~Toublan and J.~J.~Verbaarschot for 
fruitful collaboration. This work was supported in part by 
US DOE grant DE-FG-88ER40388. A different scenario for 
chiral vacuum alignment in the CFL phase was discussed 
by Lee\cite{Lee:2001jx}.


%

\end{document}